\documentclass[12pt]{article}
\usepackage[cp1251]{inputenc}
\usepackage[english]{babel}

\usepackage{graphicx}
\usepackage{amsfonts}
\usepackage{amssymb}
\usepackage{amsmath}
\usepackage[colorlinks,urlcolor=blue]{hyperref}

\usepackage{mathrsfs}
\usepackage{listings}

\begin{document}

\title{\textbf{Orbitron. Part II. Magnetic levitation}}

\author{Stanislav S. Zub$^{*}$}
\date{}
\maketitle

\tableofcontents

\newpage

\begin{abstract}
This paper devoted to proof the existence of stable quasi-periodic motions
of the magnetic dipole that is under the action of the external magnetic field and homogeneous field of gravity.
For proof this we used the group-theoretic methods of Hamiltonian mechanics, viz energy-momentum method.
Numerical simulation shows the possibility of realization of stable motions
with physically reasonable parameters of the system.
\end{abstract}

\makeatletter

\let\oldthefootnote\thefootnote
\renewcommand{\thefootnote}{\fnsymbol{footnote}}
\footnotetext[1]{Faculty of Cybernetics. Taras
Shevchenko National University of Kyiv. \\
Glushkov boul., 2, corps 6. UA-03680, Ukraine. E-mail: \url{stah@univ.kiev.ua}}
\let\thefootnote\oldthefootnote

\makeatother

%
\inputencoding{cp1251}
\newcommand{\s}[1]{\ensuremath{\boldsymbol{\sigma}_{#1}}}
\newcommand{\bsym}[1]{\ensuremath{\boldsymbol{#1}}}
\newcommand{\w}{\ensuremath{\boldsymbol{\wedge}}}
\newcommand{\lc}{\ensuremath{\boldsymbol{\rfloor}}}
\newcommand{\rc}{\ensuremath{\boldsymbol{\lfloor}}}

\thispagestyle{empty}

\section{Introduction}
\label{introduction}

\bigskip
In the paper \cite{ZubOrbitron:GOPM13} we proved that quasi-periodic motion of the magnetic dipole
in specially constructed system is exists (i.e. without gravity force).
This system is called Orbitron, and it was choosen so, to be the most simplest and the optimal
for solely magnetic task.

As it was shown in further study just that very feature of the Orbitron is not suitable
for the realization of levitation, because exactly for this system it is difficult to reach
compensate the gravity force by the use of magnetic force while ensuring the stability conditions.
The problem can be solved by combining of the magnetic field of Orbitron with the magnetic field
of the simplest type for the compensation of gravity force.

So, the system that was proposed for proof,
consists of two magnetic poles that equal by value and opposite to the sign $\pm \kappa$
that are located on the $\mp h$ of axis $z$ analogically to Orbitron.
Supplementary magnetic field for compensation is axially symmetric
(respect to the axis $z$) and  linearly depends on spatial coordinates.

We can distinguish three levels of consideration of the problem.

The first level can be called theoretical. On this level we must prove, in principal,
that stability of motion in the system does not contradict the physical laws
of the rigid body mechanics and magnetostatics.

The second level can be called physical.
At this level of consideration we must sure that the effect of stability
occurs at reasonable parameters of the system that can be provided, for example,
by existing properties of the materials.

The third level can be called as the level of experimental design.
At this level we must offer a specific scheme of the experiment
that demonstrates the corresponding example of stable motions.

Our paper involves the investigation only the first two levels.
Here it is necessary to mention the works of V. Kozorez,
who has demonstrated of the quasi-periodic motions of a small permanent magnet in a magnetic field.
In that experiment the magnet performs more then one thousands of revolutions
(that is about 6 minutes of flight).
These successful launches were to rare, because they were not based on an adequate mathematical model,
and successful initial conditions in the experiment were reproduced accidentally.

It should be mentioned that neither natural experiment nor numerical experiment
does not provide evidence of the system stability, but can provide the arguments in favor of the latter.

Therefore, our task is just to give a mathematically rigorous proof of the system stability
based on the generally accepted theory
\cite{Marsden:LM92,OrtRat:JGP99,ZubDumbStab:IntArch12,ZubOrbitron:GOPM13,GOZOrbitron:GMD12}.

\section{Mathematical model }
\label{Hamilton}

\bigskip
Our model consists of the next elements.

1. Magnetic field of the Orbitron.

There are two unlike magnetic poles $\pm \kappa$ on the axis $z$
that are placed at distance $\mp h$.
So, the magnetic field of the Orbitron $\bsym{B}^O$ has form:
\[
   \bsym{B}^O(\bsym{r})
 = \sum_{\varepsilon=\pm 1} \bsym{B}_{\varepsilon}(\bsym{r}), \qquad
   \bsym{B}_{\varepsilon}
 = -\frac{\mu_0}{4\pi}\varepsilon\kappa
   \frac{\bsym{r}-\varepsilon h \bsym{e}_z}{|\bsym{r}-\varepsilon h \bsym{e}_z|^3},
\leqno(1)\]

By its structure $\bsym{B}^O$ field has axial symmetry with respect to the axis $z$.

2. $\bsym{B}^L$ the field for the compensation with axial symmetry
that is linearly depends on the coordinates.

One can use the result of the \cite{SimHefGeimDiamagLev:AJP01}
(where decomposition is given in more general form)
\[
\begin{cases}
   B^L_z = B_0+B'z, \\
   B^L_x = -\frac12 B'x, \\
   B^L_y = -\frac12 B'y
\end{cases}
\leqno(2)\]
then in the cylindrical system we have
\[
\begin{cases}
   B^L_z = B_0+B'z, \\
   B^L_r = -\frac12 B'r.
\end{cases}
\leqno(3)\]

{\it Remark.} The total magnetic field $\bsym{B}(\bsym{x})$
that acts on the dipole is the sum of these fields, i.e.
\[ \bsym{B}(\bsym{x}) = \bsym{B}^O(\bsym{x}) + \bsym{B}^L(\bsym{x}).
\leqno(4)\]

3. Let's suggest that magnetic dipole is a small rigid body with axial symmetry
(or symmetric top) and its magnetic moment also directed along the axis of symmetry,
i.e. symmetries of the magnetic and mass distributions coincide.

Generally accepted that the configuration space of the rigid body is group $SE(3)$.
The relevant Hamiltonian formalism can be based on the classical symplectic structure
that exists on the cotangent bundle $T^*(SE(3))$
\cite{Marsden:IMS94,ZubLie:GOPM13,GOZOrbitron:GMD12}.

{\it Remark.} Most of the relationships that is required for group-theoretic description
of the Hamiltonian formalism on the $T^*(SE(3))$ can be easily derived from generalized
description, that given in \cite{ZubLie:GOPM13}.

In this study, mostly we rely on results of the energy-momentum method
that were designed for the Orbitron in \cite{GOZOrbitron:GMD12}.
However, as opposed to the mentioned work we use the inertial frame
but not the body frame that associated with the body.

So, as it shown in \cite{ZubRepers:KyivUnivCyb13} the condition of the magnetic dipole
in the inertial frame can be describe by four values ((\bsym{x}, A), (\bsym{p},\bsym{\pi})),
where
$x_i$ -- the coordinates of the mass center of the rigid body
$A$ -- the matrix of rotation for transformation from
space coordinates (in inertial frame) to the coordinates that connected
with body (in body coordinates),
$p_i$ -- the components of momentum of the body,
$\pi_i$ -- components of the angular momentum in the inertial frame.

Symplectic and Poisson structures on the $T^*(SE(3))$ are completely equivalent
\cite{Marsden:IMS94,ZubLie:GOPM13}.
Let's give the appropriate Poisson brackets in the inertial frame
\[
\begin{cases}
   \{x_i, x_j\} = 0,
   \{x_i, p_j\} = \delta_{ij},
   \{x_i, A_{jk}\} = 0,
   \{x_i, \pi_j\} = 0, \\
   \{p_i, p_j\} = 0,
   \{p_i, A_{jk}\} = 0,
   \{p_i, \pi_j\} = 0, \\
   \{\pi_i, A_{jk}\} = \varepsilon_{ijl}A_{lk},
   \{\pi_i, \pi_j\} = \varepsilon_{ijl}\pi_l,
\end{cases}
\leqno(5)\]

4. Hamiltonian system is given by expression
\[ h = T \left(\bsym{\pi},\mathbf{p}\right)
   + V\left(\bsym{\nu}, \mathbf{x} \right)
\leqno(6)\]
\[
= \frac{1}{2M}\bsym{p}^2+\frac12\alpha\bsym{\pi}^2
   + \frac12\beta\langle \bsym{\pi}, \bsym{\nu}\rangle^2
   -m \langle \mathbf{B}(\mathbf{x}), \bsym{\nu}\rangle + Mg \mathbf{x} \cdot \mathbf{e_3},
\]
where
$M$ --- the dipole mass,
$g$ --- free fall acceleration,
$m$ --- the quantity of the magnetic moment,
$\alpha=1/I_{\bot}$,$I_{\bot}=I_1=I_2$ --- moments of inertia of a symmetric top,
$\beta = \frac{1}{I_3} - \frac{1}{I_1}$,
$\bsym{\nu} = A\bsym{e_3}$, $\bsym{e}_3=(0,0,1)$ arithmetic vector (column),
and thus, $\bsym{\nu}$ --- this is the third column of the matrix $A$.

{\it Remark.}
In the Poisson reduction of the symmetric top \cite{ZubRepers:KyivUnivCyb13}
the components of 1st and 2nd columns of the matrix of rotation $A$ disappear from our description.
At that time the term of the Hamiltonian
$\beta\langle \bsym{\pi}, \bsym{\nu}\rangle^2$
can be discarded because it is the Casimir function of our Poisson structure.

However, it is does not give any advantages for the application of energy-momentum method,
so we do not spend reduction and thus we stay in the framework of the classical Poisson structure (5).
Then the value of $\langle \bsym{\pi}, \bsym{\nu}\rangle$ is an integral of motion,
but not a Casimir function, and therefore it cannot be discarded.

\section{The toral action and the momentum map }
\label{Model}

\bigskip
There are two symmetries in our case.
The first of them is associated with axial symmetry of the magnetic field and the second one,
in fact, associated with symmetry of the magnetic body.

Let's formalize two symmetries that were described above in the form of toral action:
\[ \Phi:
\begin{cases}
        (\mathbb{T}^2=S^1\times S^1)\times SE(3) \longrightarrow SE(3) \\
        (\left(R(\phi),R(\psi)\right),(\mathbf{x},A))\longmapsto
              (R(\phi)\mathbf{x},R(\phi)A R(-\psi)),
\end{cases}
\leqno(7)\]
where the matrix $R(\theta)$ is the rotation matrix around the 3rd axis at the angle $\theta$,
in particular

\[ R(\theta)\bsym{e}_3 = \bsym{e}_3
\leqno(8)\]
The action (7) on the group $SE(3)$ as on the configuration space continues (Cotangent Lift)
by the action of the same group (of torus) on the cotangent bundle $T^*(SE(3))$ as follows:
\[ \Phi_{\left(R(\phi),R(\psi)\right)}\left((\mathbf{x},A),(\mathbf{p},\boldsymbol{\pi})\right)
       = \left(R(\phi)\mathbf{x},(R(\phi)A R(-\psi)),(R(\phi)\mathbf{p},R(\phi)\boldsymbol{\pi})\right).
\leqno(9)\]
The momentum map that corresponds to this action is two-component function
on the phase space of the group, i.e. $T^*(SE(3))$
\[ \mathbb{J}((\mathbf{x},A),(\mathbf{p},\boldsymbol{\pi}))
       = (\langle\bsym{j},\bsym{e}_3\rangle,-\langle\bsym{\pi},A \bsym{e}_3\rangle),
\]
where $\bsym{j}$ --- total angular momentum of the body, or
\[ \mathbb{J}((\mathbf{x},A),(\mathbf{p},\boldsymbol{\pi}))
       = (\langle\bsym{\pi + x\times p},\bsym{e}_3\rangle,-\langle\bsym{\pi},\bsym{\nu}\rangle).
\leqno(10)\]
It is easy to check that the Hamiltonian (6) is the invariant value relative to the action (9),
using in particular (8). From that follows that components of the moment (10) are the motion integrals.

\section{The energy-momentum method }
\label{Stable}
\bigskip

This method is the most transparent way to study the relative equilibria
\cite{Arnold:Nauka89,Marsden:IMS94,GOZOrbitron:GMD12}
of Hamiltonian system with symmetry and allow both to formalize the searching
of the relative equilibria but also investigate their stability.

Energy-momentum method inherits the idea of Lagrange of the finding the conditional extremum,
but at the same time the algorithm become more simple because of using group-theoretic approaches.
The most remarkable simplification of research is that it comes down to the analyze
of some structures of linear algebra  in a single point of the relative equilibrium.

{\it Remark.} Since the research is conducted only at one point then in our case
the strength of the magnetic field at that point, the component of the Jacobian
and the Hessian are the independent variables and in particular this allowed
to divide our task into relatively independent parts.

We expect that the relative equilibrium that were found based on the energy-momentum method
can be stable with relation to some parameters of mentioned above elements of the system.

As well as for the Lagrange method the objective function has been constructing
and it given by the expression:
\[ h^{\xi} := h - \bsym{J}^{\xi} = h - \xi_1 \bsym{J}^1 - \xi_2 \bsym{J}^2
\leqno(11)\]
\[ = \frac{1}{2M}\bsym{p}^2+\frac12\alpha\bsym{\pi}^2
   + \frac12\beta\langle \bsym{\pi}, \bsym{\nu}\rangle^2
\]
\[
-m \langle \mathbf{B}(\mathbf{x}), \bsym{\nu}\rangle + Mg \mathbf{x} \cdot \mathbf{e_3}
\]
\[
-\xi_1\langle \mathbf{\bsym{\pi} + \bsym{x}\times\bsym{p}}, \mathbf{e_3} \rangle
+ \xi_2\langle \bsym{\pi}, \bsym{\nu}\rangle.
\]
Here $\xi_1$ and $\xi_2$ are the Lagrangian coefficients that have dimensions of frequency.

As will be shown soon in the relative equilibrium the magnetic dipole describes
a circle in the $xy$ plane with frequency $\xi_1$.
Interpretation of frequency $\xi_2$ somewhat more complicated but it associates
with the intrinsic angular velocity of rotation of the body.

As well as in Lagrange method the first variation of the objective function (11) determines
the relative equilibrium in the task. Also, will call  the conditions of the
fulfillment of the $\delta h^{\xi} = 0$ as necessary conditions of equilibrium.
Sufficient conditions is obtained from the analysis of the second variation $\delta^2 h^{\xi}$
\cite{Arnold:Nauka89,GOZOrbitron:GMD12}.
This analysis is the multi-step, and it is this will be the content of this article.

\section{Necessary conditions of the relative equilibrium }
\label{Stable}
\bigskip

Let $v\in T_{((\bsym{x}, A), (\bsym{p},\bsym{\pi}))}(T^*(SE(3)))$, i.e.
in representation of the right trivialization it is corresponds to
the inertial frame \cite{AbrMarsd:Found78,ZubLie:GOPM13,GOZOrbitron:GMD12},
and has form: (($\bsym{\delta x},\bsym{\delta A}),(\bsym{\delta p},\bsym{\delta \pi}))$.
Then for variations (i.e. for derivatives in the direction $v$, see \cite{GOZOrbitron:GMD12})
that constitute $h^{\xi}$ we have
\[ \delta_v T \left(\bsym{\pi},\mathbf{p}\right)
   = \mathbf{d} T \left(\bsym{\pi},\mathbf{p}\right) \cdot v
\]
\[ = \alpha\langle \bsym{\pi},\bsym{\delta}\bsym{\pi} \rangle
   + \frac{1}{M} \langle \mathbf{p}, \bsym{\delta}\mathbf{p} \rangle
  + \beta\langle\bsym{\pi}, \bsym{\nu}\rangle
     (\langle\bsym{\nu}, \bsym{\delta}\bsym{\pi}\rangle
  + \langle \bsym{\nu}\times\bsym{\pi},\bsym{\delta A}  \rangle)
\]
\[\delta_v V\left(A, \mathbf{x} \right) =  \mathbf{d} V \left(A, \mathbf{x} \right) \cdot v
\]
\[ =  -m \langle \mathbf{DB}(\mathbf{x})[\bsym{\nu}], \bsym{\delta}\mathbf{x}\rangle
   + M g \langle\mathbf{e_3}, \bsym{\delta}\mathbf{x}\rangle
   -m\langle\bsym{\nu}\times \mathbf{B} (\mathbf{x} ), \bsym{\delta A} \rangle
\]
\[ \delta_v J^{\bsym{\xi}}\left(\mathbf{x},A,\mathbf{p},\bsym{\pi}\right)
   = \mathbf{d} J^{\bsym{\xi}}\left(\mathbf{x},A,\mathbf{p},\bsym{\pi}\right)\cdot v
\]
\[ =\xi_1(\langle\bsym{e_3}, \bsym{\delta}\bsym{\pi}\rangle
     + \langle\bsym{p}\times\bsym{e_3},  \bsym{\delta}\bsym{x}\rangle
     - \langle \bsym{x}\times\bsym{e_3}, \bsym{\delta}\bsym{p}\rangle)
     - \xi_2(\langle \bsym{\nu}, \bsym{\delta}\bsym{\pi}\rangle
     + \langle \bsym{\nu}\times\bsym{\pi}, \bsym{\delta A} \rangle)
\]
i.e.
\[
\begin{cases}
   \tilde{\delta}_v T \left(\bsym{\pi},\mathbf{p}\right)
   = \alpha\langle \bsym{\pi},\bsym{\delta}\bsym{\pi} \rangle
   + \frac{1}{M} \langle \mathbf{p}, \bsym{\delta}\mathbf{p} \rangle;\\
   \delta_v V\left(A, \mathbf{x} \right)
   =  -m \langle \mathbf{DB}(\mathbf{x})[\bsym{\nu}], \bsym{\delta}\mathbf{x}\rangle
   + M g \langle\mathbf{e_3}, \bsym{\delta}\mathbf{x}\rangle
   -m\langle\bsym{\nu}\times \mathbf{B} (\mathbf{x} ), \bsym{\delta A} \rangle;\\
   \tilde{\delta_v} J^{\bsym{\xi}}\left(\mathbf{x},\bsym{\nu},\mathbf{p},\bsym{\pi}\right)
   = \bsym{d} J^{\bsym{\xi}}\left(\mathbf{x},\bsym{\nu},\mathbf{p},\bsym{\pi}\right)\cdot v \\
   = \xi_1[\langle\bsym{e_3}, \bsym{\delta}\bsym{\pi}\rangle
   + \langle\bsym{p}\times\bsym{e_3},  \bsym{\delta}\bsym{x}\rangle
   - \langle \bsym{x}\times\bsym{e_3}, \bsym{\delta}\bsym{p}\rangle]
   - \tilde{\xi}_2[\langle \bsym{\nu}, \bsym{\delta}\bsym{\pi}\rangle
    + \langle \bsym{\nu}\times\bsym{\pi}, \bsym{\delta A} \rangle]
\end{cases}
\leqno(12)\]
where the wave-symbol means the regrouping the terms in the first variation of
the associated Hamiltonian --- the summand with $\beta$ is taken from first row
and is placed into the third row (see (13)), of course, this does not affect
on the procedure of the computation and also on it result.
\[
\begin{cases}
   \bsym{\nu} = A\bsym{e_3}; \\
   \tilde{\xi}_2 = \xi_2+\beta\langle\bsym{\pi},\bsym{\nu}\rangle.
\end{cases}
\leqno(13)\]

From formula (12), collecting the coefficients of the independent variations,
we get the following necessary conditions:

\[
\bsym{\delta A}:\qquad
A\bsym{e_3}\times
[\xi_2+\beta \langle\bsym{\pi},A\bsym{e_3}\rangle) \bsym{\pi}
- m\bsym{B(\bsym{x})}]=0
\]

\[
\bsym{\delta}\bsym{x}:\qquad
\xi_1\cdot(\bsym{e_3}\times\bsym{p})
-m\bsym{DB}(\bsym{x})^T[A\bsym{e_3}]
+Mg \bsym{e_3}=0
\]

\[
\bsym{\delta}\bsym{\pi}:\qquad
\alpha\bsym{\pi}
-\xi_1 \bsym{e_3}
+(\xi_2+\beta\langle\bsym{\pi},A\bsym{e_3}\rangle)A\bsym{e_3} =0
\]

\[
\bsym{\delta}\bsym{p}:\qquad
\frac{1}{M}\bsym{p}
-\xi_1(\bsym{e_3}\times\bsym{x}) =0
\]

See also \cite{GOZOrbitron:GMD12}, but remember that in this article
was used coordinate system that associates with body.
In view of (13) the necessary conditions of equilibrium take the form:
\[
\begin{cases}
   \frac{1}{M}\bsym{p}-\xi_1(\bsym{e_3}\times\bsym{x})=0; \\
   \xi_1(\bsym{e_3}\times\bsym{p})-m\bsym{DB}(\bsym{x})^T[\bsym{\nu}]+Mg \bsym{e_3}=0; \\
   \alpha\bsym{\pi}-\xi_1 \bsym{e_3}+\tilde{\xi}_2\bsym{\nu}=0; \\
   \bsym{\nu}\times(\tilde{\xi}_2 \bsym{\pi} - m\bsym{B(\bsym{x})})=0.
\end{cases}
\leqno(14)\]

In conformity with the Lagrange approach the equations are determine
not only the dynamic variables in the equilibrium, but also
the Lagrange multipliers $\xi_1$ and $\tilde{\xi}_2$.

{\it Remark.} In the purely magnetic case \cite{ZubOrbitron:GOPM13,GOZOrbitron:GMD12}
for all relative equilibria that were investigated, the constant $\xi_2$ remains free.
Some restrictions on this value appeared only at the stage of studying the stability
of the relative equilibria.

The first two equations (14) do not contain any variables that characterize the proper rotation
of the body, nor the values of the magnetic field (only the components of the Jacobian).
The Lagrange multiplier $\xi_1$ is determined only from these equations.
First equation in (14) can be written as $\dot{\bsym{x}} = \xi_1(\bsym{e}_3\times\bsym{x})$.
This means that vector $\bsym{x}$ rotates around the axis $\bsym{e}_3$ with angular velocity $\xi_1$.
Accordingly, the vector $\bsym{p}=M\dot{\bsym{x}}$ is also rotated around this axis
with the same angular velocity.

{\it Remark.} Thus we have shown that the magnetic dipole is moved around the circle in the plane
orthogonal to the axis $z$.
{\it Farther on the search of the relative equilibria we will restrict by the condition of} $z = 0$.

By combining of first two equations of (14) we write the conditions of the relative equilibrium
in the form:
\[  m\bsym{DB}(\bsym{x})^T[\bsym{\nu}] = -M\xi_1^2\bsym{x}_\perp + Mg \bsym{e}_3.
\leqno(15)\]
Equation (15) and condition $\bsym{\nu}^2=1$ is fully define the components of
$\bsym{\nu}$ and the value $\xi_1$ independently of the other variables of the system.

Let's describe the relative positions for physical vectors taking into consideration (14)
and axial symmetry of the magnetic field.
First of all, from the axial symmetry of the field we have
$\bsym{B} \in  {\rm span}\left\{\bsym{e}_3, \bsym{x}\right\}$. From third equation
\[ \bsym{\pi} = \frac{\xi_1}{\alpha}\bsym{e}_3 - \frac{\tilde{\xi_2}}{\alpha}\bsym{\nu} \longrightarrow
   \bsym{\pi} \in  {\rm span}\left\{\bsym{e}_3, \bsym{\nu}\right\},
\leqno(16)\]
and from the fourth we have
\[ \bsym{\nu}\parallel(\tilde{\xi}_2 \bsym{\pi} - m\bsym{B(\bsym{x})}) \longrightarrow
   \tilde{\xi}_2\bsym{\pi} - m\bsym{B(\bsym{x})} = k\bsym{\nu}.
\leqno(17)\]
Substituting in (17) the expression for $\bsym{\pi}$ from (16) we obtain
\[ m\bsym{B(\bsym{x})} = \frac{\xi_1 \tilde{\xi}_2}{\alpha}\bsym{e}_3
  - \left(k + \frac{\tilde{\xi_2}}{\alpha}\right)\bsym{\nu} \longrightarrow
    \bsym{B} \in  {\rm span}\left\{\bsym{e}_3, \bsym{\nu}\right\}
\leqno(18)\]
Note that the 2nd term in this case can not be zero because the total magnetic field (4)
in the plane $z = 0$ will be with the orthogonal component $\bsym{e}_3$.
Then from aforesaid it follows that
\[ \begin{cases}
      \bsym{B} \in  {\rm span}\left\{\bsym{e}_3, \bsym{x}\right\}, \\
      \bsym{\nu} \in  {\rm span}\left\{\bsym{e}_3, \bsym{x}\right\}, \\
      \bsym{\pi} \in  {\rm span}\left\{\bsym{e}_3, \bsym{x}\right\}
   \end{cases}
\leqno(19)\]

This means that in the relative equilibrium all physical vectors of the task
except $\bsym{p}$ lie in a plane of $\{\bsym{e}_3, \bsym{x}\}$.

As it was already mentioned, the advantage of the energy-momentum method of study
of stability of the relative equilibrium that it can be reduce to the analyses
of the system of equations in any point of the relative equilibrium.
Having relation (19) we fix the plane of physical vectors $zx$.
Then, in particular
\[ \begin{cases}
      \bsym{x}_0 = r_0 \bsym{e}_1, \\
      \bsym{p}_0 = p_0 \bsym{e}_2 = M\xi_1 r_0 \bsym{e}_2, \\
      \bsym{\nu}_0 = \nu^1 \bsym{e}_1 + \nu^3 \bsym{e}_3, \\
      \bsym{\pi}_0 = \pi^1 \bsym{e}_1 + \pi^3 \bsym{e}_3
   \end{cases}
\leqno(20)\]
For a given $\xi_1$ this allows us write the solution
of the last two equations (14) as follows:
\[
\begin{cases}
   \pi_1 = \frac{m(\nu_3 B_1 - \nu_1 B_3)}{\xi_1}, \\
   \pi_3 = \frac{\xi_1}{\alpha}
         + \frac{\nu_3}{\nu_1}\frac{m(\nu_3 B_1 - \nu_1 B_3)}{\xi_1},\\
   \tilde{\xi}_2 = -\frac{\alpha m(\nu_3 B_1 - \nu_1 B_3)}{\xi_1\nu_1}.
\end{cases}
\leqno(21)\]

In this case, as it was noted above (see (15)) values of $\xi_1$, $\nu_1$, $\nu_3$
are completely independent of the $\pi_1,\pi_3,\tilde{\xi}_2$.

\section{Determination of dynamic variables in the relative equilibrium }
\label{Stable}
\bigskip

In the previous section we have formulated the equation (15) that together
with the condition $\bsym{\nu}^2=1$  allows you to find not yet certain values
$\nu_1$, $\nu_3$ and the Lagrange multiplier $\xi_1$.

These equations with regard (20) can be presented as the system:
\[
\begin{cases}
   \nu^1 B_{r,r} + \nu^3 B_{r,z} = -\frac{M}{m}\xi_1^2 r_0; \\
   \nu^1 B_{z,r} + \nu^3 B_{z,z} = \frac{M}{m}g; \\
   \nu_1^2  + \nu_3^2 = 1
\end{cases}
\leqno(22)\]
Note that in consequence of the Maxwell equations for the constant external field
without source of field in the area of dipole motion the following relationships
are performed (see in the cylindrical coordinates):
\[
\begin{cases}
   B_{r,z} = B_{z,r}, \\
   B_{r,r} + B_{z,z} = -\frac{B_r}{r_0}.
\end{cases}
\leqno(23)\]

{\it Remark.} Note that in $z =0$ plane the Orbitron's field by reason of mirror symmetry
has only vertical component and $B^O_{z,z}=0$, thus $B_{z,z} = B'$ (see (1,2)).

Work in
\[
\begin{cases}
   \zeta^2 = \frac{r\xi_1^2}{g}, \\
   \lambda = \frac{m B'}{M g}, \\
   \sigma = -\frac{B_{r,z}}{B'}
\end{cases}
\leqno(24)\]

Let's present the matrix (22) in the form
\[
    \begin{bmatrix}
       B_{r,z} & B_{z,r} \\
       B_{r,z} & B_{z,z}
    \end{bmatrix} =
    B'\begin{bmatrix}
        -\frac12  & -\sigma \\
        -\sigma & 1
      \end{bmatrix}.
\leqno(25)\]
Then (22) can be written as
\[ \begin{bmatrix}
        -\frac12  & -\sigma \\
        -\sigma & 1
   \end{bmatrix}
   \begin{bmatrix}
       \nu^1 \\
       \nu^3 \\
   \end{bmatrix}  =
   \frac1{\lambda}\begin{bmatrix}
                    -\zeta^2 \\
                     1      \\
                 \end{bmatrix}
\leqno(26)\]
and solution in the form
\[
\begin{cases}
   \nu^1 = \frac1{\lambda}\frac{\zeta^2 -\sigma}{\sigma^2 + \frac12}, \\
   \nu^3 = \frac1{\lambda}\frac{\sigma\zeta^2 + \frac12}{\sigma^2 + \frac12}.
\end{cases}
\leqno(27)\]
Quadratic equation for $\zeta^2$ is obtained, as before,
from the third equation (22) and has the following {\it proper} solution
for the case of $\sigma>0$
\[ \zeta^2 = \frac{\sigma(1 + \varrho)}{1 + \sigma^2}
           + \frac{\sigma^2 - \varrho}{1 + \sigma^2}
             \sqrt{\lambda^2 (1 + \sigma^2) - 1}.
\leqno(28)\]

The relations (20,24,27,28,21) give us the value of the all significant
physical quantities in the chosen fixed point of the relative equilibrium.

\section{The second variation of the associated Hamiltonian }
\label{Stable}

\bigskip
As mentioned above, the research of stability is based on the analysis of the second variation
of the associated Hamiltonian that is conducted in several stages. On the first step we calculate
the second variation.

\subsection{Calculating of the second variation of the potential energy }
\label{RE}

We have (see (12))
\[ \delta_v V\left(A, \mathbf{x} \right)
   =  -m \langle \mathbf{D B}^T(\mathbf{x})[\bsym{\nu}], \bsym{\delta}\mathbf{x}\rangle
   + M g \langle\mathbf{e_3}, \bsym{\delta}\mathbf{x}\rangle
   -m\langle\bsym{\nu}\times \mathbf{B} (\mathbf{x} ), \bsym{\delta A}\rangle;\\
\leqno(29)\]
\[ \delta_v\delta_v V\left(A, \mathbf{x} \right)
   = -m \delta_v[\langle \mathbf{D B}^T(\mathbf{x})[\bsym{\nu}], \bsym{\delta}\mathbf{x}\rangle
   + \langle\bsym{\nu}\times \mathbf{B} (\mathbf{x} ), \bsym{\delta A}\rangle];
\leqno(30)\]

It is evident that tensor notation will be the most suitable mathematical apparatus
for the analysis of the variations
\[ \langle \mathbf{D B}^T(\mathbf{x})[\bsym{\nu}], \bsym{\delta}\mathbf{x}\rangle
   = \nu_i B_{i,k}\delta x^k
\leqno(31)\]
\[  \langle\bsym{\nu}\times \mathbf{B} (\mathbf{x} ), \bsym{\delta A}\rangle
  = \langle \mathbf{B}(\mathbf{x} ), \bsym{\delta A}\times\bsym{\nu}\rangle
  = \nu_i\varepsilon_{iks}B_{k}\bsym{\delta A}_s,
\leqno(32)\]
where $\varepsilon$ --- the Levi-Civita symbol.

Then
\[ \delta_v\langle \mathbf{D B}^T(\mathbf{x})[\bsym{\nu}], \bsym{\delta}\mathbf{x}\rangle
   = B_{k,r}\delta\nu_k \delta x^r + B_{i,kl}\delta x^k\delta x^l
\]
\[ = B_{k,r}(\varepsilon_{ksi}\bsym{\delta A}_s\nu_i) \delta x^r
   + \nu_i B_{i,kl}\delta x^k\delta x^l
\]
\[ = \nu_i \varepsilon_{iks} B_{k,r}\delta x^r \bsym{\delta A}_s
   + \nu_i B_{i,kl}\delta x^k\delta x^l
\]
i.e.
\[ \delta_v\langle \mathbf{D B}^T(\mathbf{x})[\bsym{\nu}], \bsym{\delta}\mathbf{x}\rangle
 = \nu_i \varepsilon_{iks} B_{k,r}\delta x^r \bsym{\delta A}_s
   + \nu_i B_{i,kl}\delta x^k\delta x^l
\leqno(33)\]

Moreover
\[ \delta_v\langle\bsym{\nu}\times \mathbf{B} (\mathbf{x} ), \bsym{\delta A}\rangle
   = \delta_v (\nu_i\varepsilon_{iks}B_{k}\bsym{\delta A}_s)
\]
\[ = \nu_i\varepsilon_{iks}B_{k,r}\delta x_r \bsym{\delta A}_s
   + \varepsilon_{iks}B_{k}\bsym{\delta A}_s\delta\nu_i
\]
\[ = \nu_i\varepsilon_{iks}B_{k,r}\delta x_r \bsym{\delta A}_s
   + \varepsilon_{iks}B_{k} \bsym{\delta A}_s\varepsilon_{irt}\bsym{\delta A}_r\nu_t
\]
\[ = \nu_i\varepsilon_{iks}B_{k,r}\delta x_r \bsym{\delta A}_s
   + (\delta_{kr}\delta_{st} - \delta_{kt}\delta_{sr})B_{k} \bsym{\delta A}_s \bsym{\delta A}_r \nu_t
\]
\[ = \nu_i\varepsilon_{iks}B_{k,r}\delta x_r \bsym{\delta A}_s
   + B_{r}\nu_s \bsym{\delta A}_r\bsym{\delta A}_s
   - \nu_k B_{k}\bsym{\delta A}_r\bsym{\delta A}_r
\]
i.e.
\[ \delta_v\langle\bsym{\nu}\times \mathbf{B} (\mathbf{x} ),\bsym{\delta A}\rangle
   = \nu_i\varepsilon_{iks}B_{k,r}\delta x_r \bsym{\delta A}_s
   + B_{r}\nu_s \bsym{\delta A}_r \bsym{\delta A}_s
   - \nu_k B_{k}\bsym{\delta A}_r \bsym{\delta A}_r
\leqno(34)\]

So
\[ \delta_v\delta_v V\left(A, \mathbf{x} \right)
   = -m \delta_v[\langle \mathbf{D B}^T(\mathbf{x})[\bsym{\nu}], \bsym{\delta}\mathbf{x}\rangle
   + \langle\bsym{\nu}\times \mathbf{B} (\mathbf{x} ), \bsym{\delta A} \rangle]
\leqno(35)\]
\[ = -m \left(2\nu_i \varepsilon_{iks} B_{k,r}\delta x_r \bsym{\delta A}_s
       + \nu_i B_{i,kl}\delta x_k\delta x_l
       + B_{r}\nu_s \bsym{\delta A}_r \bsym{\delta A}_s
       - \nu_k B_{k}\bsym{\delta A}_r \bsym{\delta A}_r\right);
\]

\subsection{Calculation of the second variation of the associated Hamiltonian }
\label{Point}

\[ \delta_v\delta_v h ^\xi  = \delta_v\delta_v T \left(\bsym{\pi},\mathbf{p}\right)
   + \delta_v\delta_v V\left(\bsym{\nu}, \mathbf{x} \right)
   - \delta_v\delta_v J^{\bsym{\xi}}\left(\mathbf{x},\bsym{\nu},\mathbf{p},\bsym{\pi}\right)
\leqno(36)\]
\[  = \delta_v\tilde{\delta}_v T \left(\bsym{\pi},\mathbf{p}\right)
   + \delta_v\delta_v V\left(\bsym{\nu}, \mathbf{x} \right)
   - \delta_v\tilde{\delta}_v J^{\bsym{\xi}}\left(\mathbf{x},\bsym{\nu},\mathbf{p},\bsym{\pi}\right)
\]

We have the relation
\[ \delta_v\bsym{\nu} = \bsym{\delta A}\times\bsym{\nu};
\leqno(37)\]

For second variation of the kinetic energy, we obtain
\[\delta_v\tilde{\delta}_v T \left(\bsym{\pi},\mathbf{p}\right)
   = \alpha\langle \bsym{\delta}\bsym{\pi},\bsym{\delta}\bsym{\pi} \rangle
   + \frac{1}{M} \langle \bsym{\delta}\mathbf{p} , \bsym{\delta}\mathbf{p} \rangle
\leqno(38)\]

The variation of the potential energy is given in (35)
\[ \delta_v\delta_v V\left(A, \mathbf{x} \right)
   = -m \delta_v[\langle \mathbf{D B}^T(\mathbf{x})[\bsym{\nu}], \bsym{\delta}\mathbf{x}\rangle
   + \langle\bsym{\nu}\times \mathbf{B} (\mathbf{x} ), \bsym{\delta A}\rangle];
\leqno(39)\]
\[ = -m \left(2\nu_i \varepsilon_{iks} B_{k,r}\delta x_r\bsym{\delta A}_s
       + \nu_i B_{i,kl}\delta x_k\delta x_l
       + B_{r}\nu_s\bsym{\delta A}_r\bsym{\delta A}_s
       - \nu_k B_{k}\bsym{\delta A}_r\bsym{\delta A}_r\right);
\]

For calculation of $\delta_v\tilde{\delta}_v J^{\bsym{\xi}}$
the following relations will be useful
\[ \delta_v[\langle\bsym{p}\times\bsym{e_3},  \bsym{\delta}\bsym{x}\rangle
           - \langle \bsym{x}\times\bsym{e_3}, \bsym{\delta}\bsym{p}\rangle]
   = \langle\bsym{\delta}\bsym{p}\times\bsym{e_3},  \bsym{\delta}\bsym{x}\rangle
           - \langle \bsym{\delta}\bsym{x}\times\bsym{e_3}, \bsym{\delta}\bsym{p}\rangle
   = 2\langle\bsym{e_3}, \bsym{\delta}\bsym{x}\times\bsym{\delta}\bsym{p}\rangle
\]
\[  \delta_v\langle\bsym{\nu}, \bsym{\delta}\bsym{\pi}\rangle
   = \langle\delta\bsym{\nu}, \bsym{\delta}\bsym{\pi}\rangle
   = \langle\bsym{\delta A}\times\bsym{\nu}, \bsym{\delta}\bsym{\pi}\rangle
   = \langle\bsym{\nu}, \bsym{\delta}\bsym{\pi}\times\bsym{\delta A}\rangle
\]
\[  \delta_v\langle\bsym{\nu}\times\bsym{\pi}, \bsym{\delta A}\rangle
  = \langle\bsym{\delta}\bsym{\nu}\times\bsym{\pi}, \bsym{\delta A}\rangle
  + \langle\bsym{\nu}\times\bsym{\delta}\bsym{\pi}, \bsym{\delta A}\rangle
\]
\[ = \langle(\bsym{\delta A}\times\bsym{\nu})\times\bsym{\pi}, \bsym{\delta A}\rangle
   + \langle\bsym{\nu}\times\bsym{\delta}\bsym{\pi}, \bsym{\delta A}\rangle
\]
\[ \langle(\bsym{\delta A}\times\bsym{\nu})\times\bsym{\pi}, \bsym{\delta A}\rangle
  = \langle\bsym{\pi}, \bsym{\delta A}\times (\bsym{\delta A}\times\bsym{\nu})\rangle
\]
\[ = \langle\bsym{\pi}, \langle\bsym{\nu}, \bsym{\delta A}\rangle \bsym{\delta A}
   - \langle\bsym{\delta A}, \bsym{\delta A}\rangle\bsym{\nu}\rangle
   = \langle\bsym{\nu}, \bsym{\delta A}\rangle \langle\bsym{\pi},\bsym{\delta A}\rangle
   - \langle\bsym{\pi},\bsym{\nu}\rangle\langle\bsym{\delta A}, \bsym{\delta A}\rangle
\]
\[ \langle\bsym{\nu}\times\bsym{\delta}\bsym{\pi}, \bsym{\delta A}\rangle
 =  \langle\bsym{\nu}, \bsym{\delta}\bsym{\pi}\times\bsym{\delta A}\rangle
\]

Thus
\[ \delta_v\tilde{\delta_v} J^{\bsym{\xi}}
   = 2\xi_1\langle\bsym{e_3}, \bsym{\delta}\bsym{x}\times\bsym{\delta}\bsym{p}\rangle
   - \delta_v\tilde{\xi}_2[\langle \bsym{\nu}, \bsym{\delta}\bsym{\pi}\rangle
    + \langle \bsym{\nu}\times\bsym{\pi}, \bsym{\delta A}\rangle]
\leqno(40)\]
\[ - \tilde{\xi}_2[2\langle\bsym{\nu}, \bsym{\delta}\bsym{\pi}\times\bsym{\delta A}\rangle
                  + \langle\bsym{\nu}, \bsym{\delta A}\rangle \langle\bsym{\pi},\bsym{\delta A}\rangle
                  - \langle\bsym{\pi},\bsym{\nu}\rangle\langle\bsym{\delta A}, \bsym{\delta A}\rangle ]
\]

The value of $\tilde{\xi_2}$ in (13) formally depends not only on
the Lagrange multiplier $\xi_2$ but also on the dynamic variables.
Therefore, it would seem it is necessary to consider the variation of this value,
but in the energy-momentum method this form of $\delta^2 h^{\xi}$ is analyzed not for all variations,
but only on such that are tangent to the submanifolds of moment level, i.e.
\[ T_{\mathbf{z} } \bsym{J} \cdot v = 0 \longrightarrow \delta_v\tilde{\xi}_2 = 0
\leqno(41)\]
Thus, in all calculations of stability conditions can be assumed that $\tilde{\xi}_2$
is constant. Then
\[ \delta_v\tilde{\delta_v} J^{\bsym{\xi}}
   = 2\xi_1\langle\bsym{e_3}, \bsym{\delta}\bsym{x}\times\bsym{\delta}\bsym{p}\rangle
\leqno(42)\]
\[
   - \tilde{\xi}_2[2\langle\bsym{\nu}, \bsym{\delta}\bsym{\pi}\times\bsym{\delta A}\rangle
                  + \langle\bsym{\nu}, \bsym{\delta A}\rangle \langle\bsym{\pi},\bsym{\delta A}\rangle
                  - \langle\bsym{\pi},\bsym{\nu}\rangle\langle\bsym{\delta A}, \bsym{\delta A}\rangle ]
\]
Substituting in (36) the expressions (38), (39) and (42) we finally obtain
\[ \delta_v\delta_v h ^\xi  = \delta_v\tilde{\delta}_v T \left(\bsym{\pi},\mathbf{p}\right)
   + \delta_v\delta_v V\left(\bsym{\nu}, \mathbf{x} \right)
   - \delta_v\tilde{\delta}_v J^{\bsym{\xi}}\left(\mathbf{x},\bsym{\nu},\mathbf{p},\bsym{\pi}\right)
\leqno(43)\]
\[
= \alpha\langle \bsym{\delta}\bsym{\pi},\bsym{\delta}\bsym{\pi} \rangle
   + \frac{1}{M} \langle \bsym{\delta}\mathbf{p} , \bsym{\delta}\mathbf{p} \rangle
\]
\[ -m \left(2\nu_i \varepsilon_{iks} B_{k,r}\delta x_r\bsym{\delta A}_s
       + \nu_i B_{i,kl}\delta x_k\delta x_l
       + B_{r}\nu_s\bsym{\delta A}_r\bsym{\delta A}_s
       - \nu_k B_{k}\bsym{\delta A}_r\bsym{\delta A}_r\right)
\]
\[ - 2\xi_1\langle\bsym{e_3}, \bsym{\delta}\bsym{x}\times\bsym{\delta}\bsym{p}\rangle
   + \tilde{\xi}_2[2\langle\bsym{\nu}, \bsym{\delta}\bsym{\pi}\times\bsym{\delta A}\rangle
                  + \langle\bsym{\nu}, \bsym{\delta A}\rangle \langle\bsym{\pi},\bsym{\delta A}\rangle
                  - \langle\bsym{\pi},\bsym{\nu}\rangle\langle\bsym{\delta A}, \bsym{\delta A}\rangle ]
\]

\section{Admissible variations }
\label{numeric}
\bigskip

Stability investigation of Hamiltonian system with symmetry in accordance with
energy-momentum method reduces to analyse of the quadratic form $\mathbb{Q}$ of variations.
The positive definiteness of $\mathbb{Q}$ will mean the stability of equilibrium.

Form $\mathbb {Q}$ is derivative of form $\delta_v\delta_v h^\xi$ (see (43))
that called the {\ it initial}. In order to obtain the values of {\it reduced}
form $\mathbb{Q}$ we must use the value of the initial form not for all possible variations
of the dynamical variables, but only for variations with constraints.

These constraints are divided into two types.

The first type of the constraints were already mentioned in (see (41)).
The variations that are satisfying to the constraints will be tangent
to submanifolds of the moment level and this is the geometric sense
of the constraints.

The second type of constraints restricts the variations space in such way
that no one of the variations of this subspace will not be tangent to the orbit
of the toral action for the selected fixed point of the relative equilibrium.
Thus, the second type of constraints will give us the variations space {\it transverse}
with respect to the space that is tangent to the orbit at that point.
Let $z_0$ --- point of the relative equilibrium, and $v_1,v_2$ --- the basis vectors that are tangent
to the orbit of the toral action at that point.
Then, ${\rm span}\{v_1,v_2\}$ --- it is the tangent to the orbit space in the point $z_0$.

Let
\[
{\rm Ker} T_{z_0} \boldsymbol{J} = W  \oplus {\rm span} \{v_1,v_2\}.
\]

Then $W$ is a subspace of the admissible variations.

{\it Remark.} In contrast to the constraints of the first type the constraints
of the second type can be chosen rather arbitrarily, but keeping only the transversality condition.

Consider the constraints of the first type
\[ \begin{cases}
      \langle\bsym{e_3}, \bsym{\delta}\bsym{\pi}\rangle
   + \langle\bsym{p}\times\bsym{e_3},  \bsym{\delta}\bsym{x}\rangle
   - \langle \bsym{x}\times\bsym{e_3}, \bsym{\delta}\bsym{p}\rangle = 0, \\
     \langle \bsym{\nu}, \bsym{\delta}\bsym{\pi}\rangle
    + \langle \bsym{\nu}\times\bsym{\pi}, \delta A\rangle = 0.
\end{cases}
\leqno(44)\]
At the point $z_0$ these relations can be written in terms of coordinates
\[ \begin{cases}
      \delta \pi_3 + p_0 \delta x_1 + r_0 \delta p_2 = 0, \\
      \nu_1 \delta \pi_1 + \nu_3 \delta \pi_3 + (\nu_3\pi_1 - \nu_1\pi_3)\delta A_2 = 0.
\end{cases}
\leqno(45)\]

Consider the constraints of the second type
\[
\begin{cases}
   \langle\bsym{\nu}, \delta A\rangle   = 0, \\
   \delta p_1 = \frac{p_0}{r_0} \delta x_2
\end{cases}
\leqno(46)\]
moreover the first constraint ``prohibits'' the rotation around the axis of the body symmetry,
and the second one around the axis $z$ symmetry of the system.

Let's write via the coordinates
\[ \begin{cases}
      \delta p_1 - \frac{p_0}{r_0} \delta x_2 = 0, \\
      \nu_1 \delta A_1 + \nu_3 \delta A_3 = 0
\end{cases}
\leqno(47)\]
The relations (45) and (47) allow us to exclude the following variations:
$\delta p_1, \delta A_3, \delta \pi_3, \delta p_2$
\[ \begin{cases}
      \delta p_1 = \frac{p_0}{r_0} \delta x_2, \\
      \delta A_3 = -\frac{\nu_1}{\nu_3} \delta A_1, \\
      \delta \pi_3 =  -\frac{\nu_1 \delta \pi_1 + (\nu_3\pi_1 - \nu_1\pi_3)\delta A_2}{\nu_3}, \\
      \delta p_2 =  \frac{\nu_1 \delta \pi_1 + (\nu_3\pi_1 - \nu_1\pi_3)\delta A_2 - p_0\nu_3\delta x_1}{r_0\nu_3}
\end{cases}
\leqno(48)\]
Note that in accordance with the formula (27) $\nu_3\neq 0$.

\section{The matrix of the reduced quadratic form }
\label{numeric}

\bigskip
Let's call via $Q$ the reduced quadratic form of the initial form $\mathbb{Q}$.
In the previous section the algorithm of the matrix calculation described in detail.
It uses the components of Jacobian and Hessian of magnetic field at the point
$\bsym{x}_0=r_0\bsym{e}_1$.

One can show that in this point the corresponding matrices have the form:
\[ DB|_{z=0} =
  \begin{bmatrix}   -\frac12 B' & 0 & B_{z,r} \\
                    0 & -\frac12 B' & 0 \\
                   B_{z,r} & 0 & B'
  \end{bmatrix}
\leqno(49)\]
and the components of Hessian
\[ D^2 B_1|_{z=0} =
  \begin{bmatrix}   0  & 0  &   B_{z,rr} \\
                    0  & 0  &  0 \\
                    B_{z,rr}  & 0  &  0 \\
  \end{bmatrix}
\leqno(50)\]
\[ D^2 B_2|_{z=0} =
  \begin{bmatrix}   0  & 0  &  0 \\
                    0  & 0  & \frac{1}{r}B_{z,r} \\
                    0  & \frac{1}{r}B_{z,r} &  0 \\
  \end{bmatrix}
\leqno(51)\]
\[ D^2 B_3|_{z=0} =
  \begin{bmatrix}   B_{z,rr} &  0  & 0 \\
                    0  &  \frac{1}{r}B_{z,r} & 0 \\
                    0  & 0  & -\left(B_{z,rr} + \frac{1}{r}B_{z,r}\right) \\
  \end{bmatrix}
\leqno(52)\]

Because of complexity of computation of the quadratic form computation was conducted in Maple.

Let adopt the following order of independent variations:
$\delta \pi_2, \delta x_2, \delta A_1, \delta \pi_1, \delta x_3, \delta x_1,\delta A_2$.
Note that the variable $\delta p_3$ give us the trivial strictly positive contribution $\delta p_3^2/M$.
Therefore, this contribution is not considered in the next steps.

The resulting matrix has a block structure:
\[ Q =
  \begin{bmatrix}   Q_1  & 0  \\
                    0  &  Q_2
  \end{bmatrix},
\leqno(53)\]
where $Q_1$ --- matrix $3\times 3$, and $Q_2$ --- matrix $4\times 4$.

Matrix $Q_1$ with the next order of the variations $\delta \pi_2, \delta x_2, \delta A_1$ have the form:
\[ Q_1 =
  \begin{bmatrix}   \alpha  & 0  &  -\frac{\tilde{\xi}_2}{\nu_3}  \\
                    0  & 3 M \xi_1^2 + \frac{m\nu_3 B_{z,r}}{r_0}  &  -\frac{m B_{z,z}}{2\nu_3} \\
                    -\frac{\tilde{\xi}_2}{\nu_3}  & -\frac{m B_{z,z}}{2\nu_3}
                    &  \frac{\langle\bsym{\nu}, m\bsym{B} - \tilde{\xi}_2\bsym{\pi} \rangle}{\nu_3^2} \\
  \end{bmatrix}
\leqno(54)\]

For matrix $Q_2$ we have
\[ Q_2 =
  \begin{bmatrix}
     \frac{1}{\nu_3^2}\left(\alpha + \frac{\nu_1^2}{M r_0^2}\right)& 0  & -\frac{2\xi_1\nu_1}{r_0\nu_3} & Q_{47} \\
     0 & m\nu_3 \left(B_{z,rr} + \frac{B_{z,r}}{r_0}\right) & -m\nu_1 B_{z,rr} & Q_{57}  \\
     -\frac{2\xi_1\nu_1}{r_0\nu_3} &  -m\nu_1 B_{z,rr}& 3 M \xi_1^2 - m\nu_3 B_{z,rr}   & Q_{67} \\
     Q_{47} & Q_{57} & Q_{67}  & Q_{77} \\
  \end{bmatrix}
\leqno(55)\]
\[
\begin{cases}
   Q_{47} = \frac{\tilde{\xi}_2}{\nu_3}
          - \xi_1\left(1 + \frac{1}{\alpha M r_0^2}\right)\left(\frac{\nu_1^2}{\nu_3^2}\right); \\
   Q_{57} = m (-\nu_3 B_{z,r} + \nu_1 B_{z,z});\\
   Q_{67} = m(\nu_1 B_{z,r} +\frac12\nu_3 B_{z,z}) + 2\frac{\xi_1^2}{r_0\alpha}\frac{\nu_1}{\nu_3};\\
   Q_{77} =  \frac{\xi_1^2}{\alpha}\left(1 + \frac{1}{\alpha M r_0^2}\right)\frac{\nu_1^2}{\nu_3^2}
          + \langle\bsym{\nu},m\bsym{B} - \tilde{\xi}_2\bsym{\pi}\rangle
\end{cases}
\leqno(56)\]

\section{The conditions of positive definiteness of the quadratic form $\mathbb{Q}$ }
\label{numeric}

\bigskip
Because of block structure of matrix $Q$  the study of the positive definiteness
can be devided on the two separate studies of the matrices $Q_1$  and $Q_2$.

Let's use the Sylvester's criterion \cite{Gelfand:LLA98}.
For this purpose we must check the determinant of the upper-left submatrices
of our matrix on the positive definiteness.

For example, for $Q_1$ it will be the next sequence of submatrices
$Q_1^{[1]},Q_1^{[2]},Q_1^{[3]}$:
\[ Q_1^{[1]} = [\alpha],\quad Q_1^{[2]} =
  \begin{bmatrix}   \alpha  & 0   \\
                    0  & 3 M \xi_1^2 + \frac{m\nu_3 B_{z,r}}{r_0} \\
  \end{bmatrix},\quad
  Q_1^{[3]} = Q_1
\leqno(57)\]
and so, the relation must be performed
\[ {\rm det}(Q_1^{[1]})>0,\quad {\rm det}(Q_1^{[2]})>0,\quad {\rm det}(Q_1)>0.
\leqno(58)\]
Accordingly for $Q_2$ we have
\[ {\rm det}(Q_2^{[1]})>0,\quad {\rm det}(Q_2^{[2]})>0,\quad {\rm det}(Q_2^{[3]})>0,\quad {\rm det}(Q_2)>0.
\leqno(59)\]

{\it Remark.} Note that more weaker, but necessary conditions of the positive definiteness of the matrix $Q$
is the positivity of its diagonal elements.

Consider the matrix $Q_1$. The first condition (58) is trivial,
since wittingly $\alpha>0$. Then the second condition can be written as
\[ 3 M \xi_1^2 + \frac{m\nu_3 B_{z,r}}{r_0} > 0.
\leqno(60)\]
The third condition can be represented in the form
\[
\left(3 M \xi_1^2 + \frac{m\nu_3 B_{z,r}}{r_0}\right)
\left(\langle\bsym{\nu}, m\bsym{B} - \xi_2\bsym{\pi} \rangle - \frac{\xi_2^2}{\alpha}\right)
- \frac14 m^2 B_{z,z}^2 > 0.
\leqno(61)\]
Second multiplier in (61) by using (21) can be simplified to
\[ \langle\bsym{\nu}, m\bsym{B} - \xi_2\bsym{\pi} \rangle - \frac{\xi_2^2}{\alpha}
   = \frac{m B_1}{\nu_1}.
\leqno(62)\]
It is obvious that from (61) with regard to (62) (originally $m > 0$) and it follows that
\[ \frac{m B_1}{\nu_1} > 0 \longrightarrow \frac{B_1}{\nu_1} > 0.
\leqno(63)\]

This condition physically means that the magnetic moment seeks to orientate oneself by the field.
It should be noted that in the supporting point of the relative equilibrium, where $z = 0$,
only field $\bsym{B}^L$ has a component of $B_1$ that is different from zero.

Then
\[ \frac{m B_1}{\nu_1} = -\frac{m B'r_0}{2\nu_1} > 0.
\leqno(64)\]

Dividing (61) on the positive value (64) and after multiplying result on $r_0$ we obtain
\[ 3 M \xi_1^2 r_0 +  m\nu_3 B_{z,r}  + \frac12 m\nu_1 B' > 0,
\leqno(65)\]
remind (see (1,2)) that for $z = 0$ we have $B_{z,z} = B'$.

The last term in (65) due to (64) is less than zero
i.e. condition (65) is a more stringent condition than (60).
Thus, the condition of (65) is the condition of the positive definiteness of the matrix $Q_1$.

Let's consider the condition of the positive definiteness of the matrix $Q_2$.

The first condition of (59) holds trivially.

It is interesting to consider the remaining diagonal elements of the matrix.

Condition $Q_{77}$ is easily derived from (62),(63).

Condition ${\rm det}(Q_2^{[2]})>0$ is reduced to the positivity of the second of diagonal element
\[  m\nu_3 \left(B_{z,rr} + \frac{B_{z,r}}{r_0}\right) = -m\nu_3 B_{z,zz} > 0.
\leqno(66)\]
This simple condition is purely geometrical and not dependent of the others.

After rather cumbersome transformations the condition ${\rm det}(Q_2^{[3]})>0$
can be written in the form
\[ 3 M \xi_1^2 - \left(\frac{1}{\nu_3^2}\frac{1 + \frac{\nu_1^2}{\alpha M r_0^2}}
                                       {1 - \frac{\nu_1^2}{3\alpha M r_0^2}}\right)
                 \left(1 + \nu_1^2\frac{B_{z,r}}{r_0 B_{z,zz}}\right)
                 m\nu_3 B_{z,rr} > 0;
\leqno(67)\]

When $\nu_1\rightarrow 0$ this condition moves to the condition of positivity of the 3rd
of diagonal element of the matrix $Q_2$.

For condition ${\rm det}(Q_2)>0$ we can get a few members of the Laurent
expansion in the small parameter $\nu_1$, but even in this approximation
the expressions are too cumbersome for the analysis in analytical representation
not to mention about exact expression for ${\rm det}(Q_2)$.
Therefore, not seems advisable to trying infer the conditions of positive definiteness
in analytical representation. Especially because the expression includes the variables
$\xi_1$, $\nu_1$, $\nu_3$, $\tilde{\xi_2}$ that are not elementary.

The main aim of this paper is to prove of stability of the relative equilibrium
of the Hamiltonian system, it mean that exists of physically stable orbital motion of the magnetic dipole
under the mutual action of axially symmetric magnetic field and gravity field (the levitation of the Orbitron).

Conditions (58) and (59) are the {\it analytic} conditions of stability of the levitating Orbitron.
Thus, we solve the task on the first (or theoretical) level. To solve the task on the second
(or physically) level it is sufficient show the reasonable parameters of the system,
for which the conditions of stability executed. Of course, if there is one such set,
then there are infinitely many such sets.

As will be shown in the next section such parameters are exist.

\section{Numerical simulation }
\label{numeric}

\bigskip
The determinants (58),(59) were programmed. Maple allows to return results
both in the symbolic form and as the numerical values of the functions.

Guided by the formulas (63),(65)--(67) and taking into account the physical arguments,
the suitable model parameters have been found.
For these parameters all condition (58),(59) are performed numerically.

Initial conditions, magnetic materials and geometrical parameters were selected.

The magnetic poles and tablets (the magnetic dipole)
were made of $Nd-Fe-B$ and have the following physical characteristics:
$\rho = 7.4\cdot 10^3(kg/m^3)$ --- density of the material and $B_r = 0.25 (T)$ --- residual induction.

Then the magnetic ``charge'' of the Orbitron's poles required to ensure the stability will be
$\kappa = 351.5625 (A \cdot m)$, and the distance between the poles is $L = 2h = 0.1 (m)$.

Linear field for compensation of the gravity force is characterized by two parameters of our model
\[
\begin{cases}
     B'  = 0.35723477320570427127 \quad (T/m), \\
     B_0 =2.985 \quad (T).
\end{cases}
\leqno(68)\]

Movable magnet (dipole) selects in the form of disk with diameter $d=0.014 (m)$
and height $l = 0.006 (m)$. Its parameters are as follows: the magnetic moment
$\mu = 0.18375 (A\cdot m^2)$, $M = 0.00683484 (kg)$ with mass and
$\alpha=1/I_{\bot} = 0.9594 (1/(kg\cdot m^2))$, where $I_{\bot} = I_1 = I_2$
the moment of inertia of the body.

For these parameters of our model in the point of
the researched relative equilibrium total magnetic field will be
\[
\begin{cases}
     B_1 =-0.17861738660285213564 \quad (T), \\
     B_3 = 2.9898002901596414059 \quad (T), \\
\end{cases}
\leqno(69)\]

So for the orbit with $r_0 =1.5(m)$ and $h = 0.075(m)$
we obtain the following parameters for the stable relative equilibrium (see also (20)):
\[
\begin{cases}
   \xi_1 = 6.6142 \quad (rad/sec), \\
   \tilde{\xi}_2 =-138.8134577 \quad (rad/sec), \\
   \nu_1 =-0.059625567564610698431, \\
   \nu_3 = 0.99822081309327449053, \\
   \pi_1 =-0.86270609223278328121 10^{-6} \quad (kg\cdot m^2/sec), \\
   \pi_3 = 0.15132393025362293319 10^{-4} \quad (kg\cdot m^2/sec).
\end{cases}
\leqno(70)\]

With these parameters of the system all the conditions of stability (58),( 59) are satisfied.
It seems that these values appear to be quite reasonable for the experimental implementation.

In addition the numerical simulation of our model conducted base on the ODE
\cite{ZubOrbitron:PoS09,ZubOrbitron:GOPM13,DullinLev:PhysD99}
that describe the dipole motion in the external magnetic and gravitational field.
\[
   \begin{cases}
      \dot{\vec{x}} = \vec{p}/M, \\
      \dot{\vec{p}} = \nabla (\vec{\mu}\cdot\vec{B}) - M g \vec{e}_3, \\
      \dot{\vec{\mu}} = (\vec{\pi}\times\vec{\mu})/I_\bot, \quad \vec{\mu}=m\vec{\nu}, \\
      \dot{\vec{\pi}} = \vec{\mu}\times\vec{B}.
\end{cases}
\leqno(71)\]

For the relative equilibrium with parameters (68) -- (70)
we calculated the sheaf of the trajectories for the movable magnetic dipole.
Random initial values were simulated in the neighborhood of the relative equilibrium
so that deviation from the relative equilibrium was less then one percent
from the corresponding value dynamical variable. Totally 100 tosses was done.
And Monte Carlo method confirms the stability of the all trajectories (by ten full turns).
Note that in the case of unstable motion of the magnetic dipole it falls down or flys away,
usually, on the first turn of the orbit.

As it was mentioned above these tests is not intended to prove system stability
because it is proved analytically and that is the main core of this work.
But very important that the test is based on (71) and completely independent
of group-theoretical methods of Hamiltonian mechanics that were applied above.

\bibliographystyle{elsarticle-num}
\bibliography{leveng-bibdb}

\end{document}